\documentstyle[aps,pra,twocolumn,epsfig]{revtex}

\title{Entanglement and Extreme Spin Squeezing}
\author{Anders S\o rensen and Klaus M\o lmer\\
Institute of Physics and Astronomy, University of Aarhus,
DK-8000 \AA rhus C., Denmark}

\begin{document}
\maketitle
\begin{abstract}
For any mean
value of a cartesian component of a spin vector we identify 
the smallest possible uncertainty in any of the orthogonal
components. The corresponding states are optimal for
spectroscopy and atomic clocks.
We show that the results for different spin $J$ can
be used to identify entanglement and to quantify the depth
of entanglement in systems with many particles. With the procedure
developed in this letter,
collective spin measurements on an ensemble  of particles can be used as
an experimental proof of multi-particle entanglement.
\end{abstract}

\bigskip

The commutator relation for angular momentum operators
leads to Heisenberg's uncertainty relation for the 
Cartesian components
\begin{eqnarray}
\Delta J_x\cdot\Delta J_y \geq |\langle J_z\rangle|/2.
\label{Heisenberg}
\end{eqnarray}
Without violating Heisenberg's uncertainty relation, it is 
possible to redistribute the uncertainty unevenly between $J_x$ and
$J_y$, so that a measurement of either $J_x$ or $J_y$ become more precise
than the standard quantum limit $\sqrt{|\langle J_z\rangle|/2}$. States
with this property are called  spin squeezed states
in analogy with the squeezed states of a harmonic oscillator.

A two-level atom can be represented as a spin 1/2 system,
and in experiments on a large number of atoms  $N$ which all start
out in the same initial state and which are all subject to the same
Hamiltonian, one can conveniently express the collective observables
of the gas by means of an effective spin $J=N/2$, so that, {\it e.g.}, the
difference in number of atoms populating the two internal states is
given by the  $J_z$-component. The state with all atoms in the 
"spin up" internal state,  is equivalent to the $|J_z=J\rangle$
eigenstate of the macroscopic spin. If one measures the $x$ component of
the spin of a single atom, it is projected onto the internal
superposition states $(|\downarrow\rangle \pm 
|\uparrow\rangle)/\sqrt{2}$ with equal probability. The value of the total
$J_x$ is given by the difference in the number of atoms in the two states,
and it fluctuates binomially with a variance $J/2$, 
which  matches precisely the
equality sign in (\ref{Heisenberg}) with $\Delta J_x = \Delta J_y$.

The states and the amount of spin squeezing produced by applying
Hamiltonians $J_x^2$ and $J_x^2-J_y^2$ to an initial 
$|J_z=J\rangle$ state have been studied \cite{ueda}, and the 
squeezed states which satisfy the equality sign in  (\ref{Heisenberg}),
the so-called minimum-uncertainty-product states,
have been identified \cite{rashid,eberly,agarwal2}. 
Interaction of atoms with broadband squeezed light
\cite{agarwal,kuzmich}, is an experimentally verified  means to
produce spin squeezing \cite{hald}.
Spin squeezing based on ideas from quantum computing
was recently suggested \cite{cat-theory}, and recent ideas for neutral atom
spin squeezing 
based on QND detection of the atomic spin state \cite{rochester} and on 
the collisional interactions between
atoms \cite{lattice,anders} give reason to believe that 
sizable spin squeezing may be much easier to achieve than optical squeezing. 

In Ramsey type spectroscopy on a collection of two-level atoms, a signal
proportional to the length of  
the  mean collective spin pointing, say, along the $z$-axis is recorded and
the noise is 
given by the uncertainty of one of the orthogonal components. 
Wineland {\it et al.} have shown \cite{Win94} that the frequency resolution
in spectroscopy on $N$ two-level atoms contains the factor 
\begin{eqnarray}
\xi=\frac{\sqrt{2J} \Delta J_x}{|\langle J_z\rangle|},
\label{winxi}
\end{eqnarray}
which is reduced by spin squeezing.
In this way, spin squeezing
becomes an important ingredient  in high precision spectroscopy 
and in  atomic
clocks, which are at present limited precisely by the fundamental spin
noise \cite{santarelli}. 
Furthermore, spin squeezing is an important ingredient in quantum
information, because the ensuing quantum entanglement
leads to possibilities, {\it e.g.} for atomic teleportation
\cite{polzik}.

In the derivation of Eq.~(\ref{winxi}) it is assumed that no other sources
of noise are present. The states which minimize the quantity $\xi$ are 
obtained in the limit $\langle J_z \rangle,\Delta J_x \rightarrow 0$, where
any  other
source of  noise will, however, deteriorate the spectroscopic resolution.  
In practice, the ideal states for 
spectroscopy are therefore states which minimize the noise $\Delta J_x$ for a
given not too small value of $\langle J_z\rangle$.
In the present Letter we identify this minimum, {\it i.e.}, we optimize the
signal to noise ratio by identifying the states with minimum quantum noise
for a given value of the signal. For photons a similar analysis has been
performed in Ref.~\cite{yuen}. Having identified
the minimum of $\Delta J_x$ we use this information to derive an
experimental criterion for entanglement. 
A measurement of two collective operators for a collection of atoms 
produces a sufficient criterion for entanglement which can even
quantify the ``depth" of entanglement, {\it i.e.}, the minimum number of
particles forming multi-particle entangled states in the sample.

\begin{figure}
  \epsfig{width=7.4cm,file=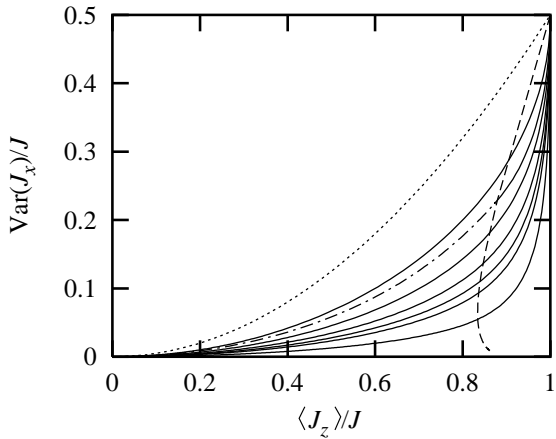}
  \caption[]{Maximal squeezing  for different values of $J$. The curves
    starting at the origin represent the minimum obtainable variance as a
    function of the mean spin. Starting from above, the curves represent
    $J=1/2,1,3/2,2,3,4,5,10$. The dotted curve for $J=1/2$ is the limit
    identified in Ref.~\cite{anders}. The full curves are obtained by
    diagonalization of the operator  $\mu J_z+J_x^2$. The dashed curve
    represents the position of a bifurcation in the solution for
    half-integer spins. To the right of this curve the diagonalization may
    be applied. To the left of the curve the minimum is found by a
    variational calculation (dash-dotted curve for $J=3/2$).}
  \label{fig:maxsqueez}
\end{figure}

To get a lower limit on  $\Delta J_x$ as a function of $\langle J_z\rangle$,
one can use the inequality $\langle J_x ^2\rangle + \langle
J_y^2\rangle +\langle 
J_z\rangle^2 \leq J(J+1)$ which together with Heisenberg's uncertainty
relation (\ref{Heisenberg}) yields the limit
\begin{eqnarray}
 (\Delta J_x)^2 &\geq& \frac{1}{2}{\Big [}J(J+1)-\langle J_z\rangle^2
 \nonumber\\
&&\quad -\sqrt{(J(J+1)-\langle J_z\rangle^2)^2-\langle J_z\rangle^2}{\Big ]}.
\label{est}
\end{eqnarray} 
This does not present a tight minimum for Var$(J_x)$,
but for large $J$ and $\langle J_z \rangle \approx J$ it is close to
the actual minimum found by the 
numerical approaches discussed below. For low values of  $\langle J_z
\rangle$ it differs by a factor of two. The precise analysis of the minimum
becomes quite different for integer spins and for half-integer spins, and
we shall deal with them separately:

For integer spins our calculations show that the state
which minimizes Var$(J_x)$ for a given $\langle J_z\rangle$
has vanishing $\langle J_x\rangle$ and $\langle J_y\rangle$, so that it is
also a
minimum of $\langle J_x^2\rangle$. Accordingly, these states can be found
by minimizing $\mu \langle J_z\rangle+\langle J_x^2\rangle$,
where $\mu$ is a Lagrange multiplier, ensuring the value of
$\langle J_z\rangle$. For $J$-values up to several hundred, it is 
straightforward to numerically determine the minimum, by determining
the smallest eigenvalue of the operator $\mu J_z+J_x^2$  for a wide range
of values of 
$\mu$. By determining $\langle J_z\rangle$ and $\langle J_x^2\rangle$ 
in the corresponding eigenstates one finds exactly the minimum value of
Var($J_x$)=$\langle J_x^2\rangle$ for the particular value of 
$\langle J_z\rangle$. The results for different values of $J$ are shown in
Fig. \ref{fig:maxsqueez}. For $J=1$ it is possible to diagonalize $\mu
J_z+J_x^2$ analytically, 
and we get Var$(J_x)_{min}=(1-\sqrt{1-\langle J_z \rangle^2})/2$. 

For half-integer spins, it is not true that the state minimizing
$\langle J_x^2\rangle$ also minimizes Var$(J_x)$ at a given value of $\langle
J_z\rangle$. The reason is that for half-integer spins, the operator
$J_x^2$ has eigenvalues $\frac{1}{4},\ \frac{9}{4}, ... $, and its
expectation value will thus always exceed $\frac{1}{4}$. The variance
of $J_x$, however, can come arbitrarily close to zero, if the system approaches
a $J_x$ eigenstate. 
Consider for instance a  $J=\frac{1}{2}$ particle, where all (pure) states
can be obtained as a simple rotation of the spin up state. In this case the
components perpendicular to the mean spin 
are never squeezed; their variances are both $\frac{1}{4}$.
But if we  compute the
variance of $J_x$ and the mean value of $J_z$, one finds that they
obey the relation, Var$(J_x)_{min} = \langle J_z\rangle^2$, where both
sides approach
zero when the state approaches either of the two $J_x=\pm \frac{1}{2}$
eigenstates. 
In that case, of course, the mean spin
also has a component along the $x$ direction. The state is spin squeezed
in the sense of the relation (\ref{Heisenberg}), but not
in the sense  where one deals explicitly with 
a spin component perpendicular to the mean spin vector. 

\begin{figure}
  \epsfig{width=7.4cm,file=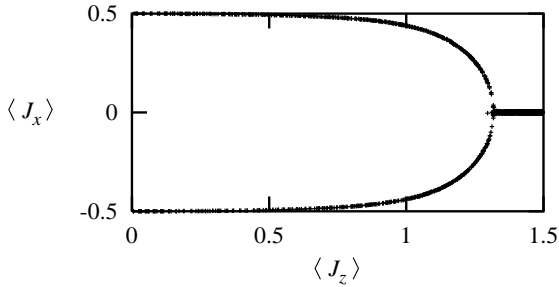}
  \caption{Bifurcation in the solution for $J=3/2$. The points represent
    the mean value of the spin in the maximally squeezed states. The points
    are obtained by a Monte Carlo variational calculation which minimizes
    $\mu \langle J_z \rangle +{\rm Var} (J_x)$. Above $\langle J_z\rangle
    \approx 1.32$ there is a unique solution with $\langle J_x\rangle
    =0$.  Below  $\langle J_z\rangle
    \approx 1.32$ the solution bifurcates, and $\langle J_x \rangle$
    approaches $\pm 1/2$.}  
  \label{fig:bifur}
\end{figure}

For large half-integer $J$ it is more difficult to find the
most squeezed states. The reason is that the problem cannot be
formulated as a linear quantum mechanics problem like
the diagonalization of an operator containing a Lagrange multiplier
term, which we used for integer spins. To compute a variance, we have to
determine the square of a  
mean value which is an expression to fourth order in wave function
amplitudes. It is easy, however, to implement a Monte Carlo variational
calculation which minimizes  $ \mu \langle J_z\rangle+{\rm Var}(J_x)$, 
by randomly modifying
the amplitudes of a state vector as long as the variational functional
is reduced. Like above, the Lagrange multiplier term is used
to adjust the mean value of $J_z$, so that for each value of $\mu$ the
identified state vector  minimizes Var$(J_x)$ for the 
given value of $\langle J_z\rangle$.  
When applied to larger half-integer values of $J$ this method shows that
large values of $\langle J_z\rangle$ are accompanied
by vanishing mean values of $J_x$ and $J_y$, and the solution thus  
coincides with the one found by the diagonalization method. 
But, for a critical value of
$\langle J_z\rangle$, the solution bifurcates,
and two states with opposite nonvanishing mean values of $J_x$ have the
smallest variance. See Fig. \ref{fig:bifur}. These states approach the two
$J_x=\pm \frac{1}{2}$ 
in the limit of small $\langle J_z\rangle$. Due to the noise in the
simulation, the Monte-Carlo method is not 
efficient for a precise determination of  the critical point of the
bifurcation for large values of $J$. Before the bifurcation the
state is the eigenstate corresponding to the lowest eigenvalue of the
operator  $\mu J_z+J_x^2$, and after the bifurcation the state is 
a superposition of the different eigenvectors with amplitudes on
states with higher eigenvalues.  We can therefore determine the position of
the bifurcation 
from the properties of the eigenvectors, and for different values of $J<
100$, 
we find that the bifurcation happens in the interval 
\begin{eqnarray}
0.83 <\frac{\langle J_z \rangle}{J}<0.88,
\label{bifu}
\end{eqnarray}
except for the special case $J=1/2$ where $\langle J_z \rangle=J$ at the
bifurcation. 
If we do not break the $\pm J_x$ symmetry the variance from this point
flattens out to the value $\frac{1}{4}$, but in either of the
states with the broken symmetry, the variance decreases
towards zero when smaller values of $\langle J_z\rangle$ and
non-vanishing values of $\langle J_x\rangle$ are considered.
The position of  the
bifurcation is plotted in Fig. \ref{fig:maxsqueez}. To the right of
the dashed line the minimum may be 
found by diagonalization. To the left of the curve the variational approach
has to be applied for half-integer spins.

It is  the
non-linearity of the problem that leads to the bifurcation and
symmetry breaking of the solution. Classical approximations to
many body quantum problems often show such bifurcations associated
with phase transitions in the problem, e.g., lasing. It is interesting
that a similar phenomenon appears here, in the study of
a single quantum system in a (very) low-dimensional Hilbert space.  
We emphazise that we are not discussing an extension of
quantum theory to include non-linear terms, we are simply identifying
the quantum states that minimize a variance, and this is a non-linear
problem. 

Since we have identified the maximally squeezed states as eigenstates of
the operator $\mu J_z+J_x^2$, even without having explicit expressions
for these eigenstates, we can devise a method to produce them.
This method only works for integer spin, and for half-integer spin
which are squeezed to values of the mean spin exceeding the
value at the bifurcation (\ref{bifu}).
The system is first prepared in the $|J_z=J\rangle$ eigenstate, and
one switches on a Hamiltonian $H(t)=\omega J_z + \chi(t) J_x^2$, where
$\chi(t)$ increases very slowly from the value zero and where
$\omega<0$. If the state of the 
spin follows this Hamiltonian adiabatically, it evolves 
through the minimum energy eigenstates of the instantaneous $H(t)$,
which is precisely the family of states identified by the above
diagonalization procedure. 
The adiabatic process may be difficult to perform in physical systems 
of interest, and for practical purposes it is relevant to point
out that the straightforward application of a Hamiltonian $H=J_x^2$,
also leads to spin squeezing \cite{ueda} and 
 in the regime with large $|\langle
J_z\rangle |$, the  squeezing resulting from this Hamiltonian is actually
close to the optimum. The 
Hamiltonian $H=J_x^2-J_y^2$, also discussed in Ref.~\cite{ueda}, 
leads to similar squeezing
for large $|\langle J_z\rangle |$, and it follows the optimum for
a larger range of parameters. 

We have identified the minimum variance of $J_x$ given the value
of $\langle J_z\rangle$. Any measurement of these two quantities can
be plotted as a point in Fig. \ref{fig:maxsqueez}, and this point must lie on
or above the   
curve for the relevant $J$. We note that the curves depend on $J$, and
in the chosen units, large spins can be more squeezed than small
spins. This  implies that the  collective spin variables 
$\vec{\cal{J}}=\sum_i \vec{J}_i$ for several spin $J$ particles 
can be more squeezed than the individual spins themselves. 
We will now show that this requires 
the state of the spins to be an entangled state. It is already known
\cite{anders}, that for spin $1/2$ particles, 
reduction of the parameter $\xi$  below unity for the
collective spin implies entanglement of the spins. We 
generalize this property to arbitrary spins.

A separable (non-entangled) state of $N$ spin $J$ particles is defined as a
weighted sum of 
products of density 
matrices with positive weights $p_k$ \cite{werner,peres}:
\begin{eqnarray}
\rho = \sum_k p_{k} \rho_1^{(k)}\otimes\rho_2^{(k)} ...
\otimes \rho_N^{(k)},
\label{separable}
\end{eqnarray}
where $\rho_i^{(k)}$ is the density matrix of the $i^{th}$ particle
in the $k^{th}$ term of the weighted sum.
The variance of ${\cal{J}}_x$ in such a state obeys the inequality
\begin{eqnarray}
\mbox{Var}({\cal{J}}_x)&\geq& \sum_k p_k \sum_{i=1}^N (\Delta J_x^2)_i^{(k)}
\nonumber \\
&\geq& 
\sum_k p_k \sum_{i=1}^N J F_J(\langle J_z\rangle_i^{(k)}/J),
\label{variance}
\end{eqnarray}
where the function $F_J(\cdot)$ is the minimum variance of $J_x$ divided by
$J$ for the
spin $J$ 
particle, i.e. the curves plotted in Fig. \ref{fig:maxsqueez}, and $\langle
J_z\rangle_i^{(k)}$ is the mean value of $J_z$ of the $i^{th}$ particle
in the $k^{th}$ realization in the sum (\ref{separable}). 

As it appears from Fig. \ref{fig:maxsqueez}, all
the curves $F_J(\cdot)$ are convex. We can prove this property 
for integer spins, and for half integer spins in the range of large
$|\langle J_z\rangle|$, by
considering  the  production of the states by adiabatic passage from the 
$|J_z=J\rangle$ eigenstate.  The positive factor
in front of the $J_x^2$ component in the Hamiltonian $\omega J_z +
\chi(t)J_x^2$ is gradually increased, and the rate of change of $\langle
J_x^2\rangle$ and of 
$\langle J_z\rangle$ at any given time are given by Ehrenfest's theorem:
\begin{eqnarray}
\frac{d}{dt}\langle J_x^2\rangle &=& \frac{1}{i\hbar}\langle 
[J_x^2,\omega J_z]\rangle\nonumber \\
\frac{d}{dt}\langle J_z\rangle& =& \frac{1}{i\hbar}\langle [J_z,
\chi(t)J_x^2]\rangle.
\label{time}
\end{eqnarray}
The mean values on the right hand side should be evaluated in the
maximally spin squeezed state, {\it i.e.}, they are not known explicitly. 
But, we observe that they contain the same operator, and the ratio
between the two rates of changes is therefore simply $-\omega/\chi(t)$. This
implies that along the family
of maximally squeezed states, the relative change of $\langle J_x^2
\rangle$ and $\langle J_z\rangle$,
{\it i.e.}, the slope of the curve $F_J(\cdot)$, is monotonically
increasing (since $\chi(t)$ is an increasing function of time and
$\omega<0$). 
It follows that the second derivative of the function $F_J(\cdot)$ is
positive, i.e., the function is convex.

From the convexity follows that the functions $F_J(\cdot)$ 
obey Jensen's inequality,
which states that any linear combination of $F_J(a_i)$'s with positive
coefficients is larger than or equal to
the function $F_J$ evaluated on the linear combination of the arguments.
It therefore follows that in any separable state
\begin{eqnarray}
\mbox{Var}({\cal{J}}_x)& \geq& \sum_k p_k NJ
F_J{\left(\frac{1}{NJ}\sum_{i=1}^N \langle 
J_z\rangle_i^{(k)}\right)} \nonumber \\
&\geq &NJ F_J{\left(\sum_k p_k \frac{1}{NJ}\sum_{i=1}^N \langle
J_z\rangle_i^{(k)}\right)} \\
\label{variancefin}
  & &\hspace{2.5cm} = NJ F_J{\left( \frac{\langle
  {\cal{J}}_z\rangle}{NJ} \right)}.\nonumber
\end{eqnarray}

This relation is a main result of this paper.
In an experiment with a collection of $N$ spin $J$ particles, it is
possible to measure the collective ${\cal{J}}_z$ and ${\cal{J}}_x$, and to
determine their mean value and variance.  If the data-point $(\langle
{\cal{J}}_z\rangle/NJ$, Var$({\cal{J}}_x)/NJ)$ lies below the
relevant curve in Fig. \ref{fig:maxsqueez}, the systems are not in a separable state,
{\it i.e.}, they are experimentally proven to be in an entangled state.

The extent to which the measured data point falls below the curve
in the plot is a measure of the degree of entanglement. 
A quantitative measure of entanglement in a multi-particle
system is the number of elements that must at least have gone together 
in entangled states.  We define a $k$ particle entangled state to be a
state of $N$ particles which {\it cannot} be decomposed into
a convex sum of products of density matrices with all density matrices 
involving less than $k$ particles: at least one of the terms
is a $k$ particle entangled density matrix.  
If, for example, $N$ spin $\frac{1}{2}$ particles form $N/2$ entangled
pairs, the degree of macroscopic spin squeezing of the system is
limited by the inequality (\ref{variancefin}) with $J=1$ and $N$ replaced
by $N/2$. 
If the measured macroscopic $\langle {\cal{J}}_z\rangle$ and
Var$({\cal{J}}_x)$ 
also lie below the corresponding $J=1$ curve, the measurement
unambiguously implies that the systems are entangled in larger than
binary ensembles.  The size of these ensembles is a measure of the
``depth" of entanglement, which can be determined experimentally. This
criterion may be compared to the one used in \cite{cat-exp} where the
fidelity of production of a maximally entangled $N$-particle states 
is used as a proof
of $N$-particle entanglement. 

As a final point we demonstrate how our procedure can be applied to identify
substantial multi-particle entanglement in recent theoretical 
proposals for spin squeezing. In Ref. \cite{anders} it is  
predicted that appreciable spin squeezing of atoms can be obtained in a
two-component Bose-Einstein condensate. In the calculation,  
a reduction of Var$({\cal{J}}_x)$ by a factor of 1000 is found for a
reduction of $\langle {\cal{J}}_z\rangle$ of only 1$\%$ with $10^5$ atoms
in the 
condensate. Using Eq.~(\ref{est}) these numbers 
imply a depth of entanglement of $\sim 2\cdot 10^4$. 
In ion traps it has been shown that it is possible to implement a
Hamiltonian $J_x^2$ by applying bichromatic light to all ions in the
trap \cite{cat-theory}. This Hamiltonian can be used to create a maximally
entangled state of all the ions. If the decoherence in
the trap is such that one cannot produce a maximally entangled state, a
different strategy could be to apply the light for a short time so that
squeezing is produced. For small times the squeezing obtained from the
Hamiltonian $J_x^2$ is close to the optimal curves in
Fig. \ref{fig:maxsqueez}, and our theory
identifies a depth of
entanglement close to the total number of ions in the trap. In this way one
could produce and verify the production of an entangled state of many ions. 


We have
considered squeezing and 
entanglement related to collective vector operators ${\cal{J}}_z$ and
${\cal{J}}_x$. 
We emphazise that the collective spin components of multi-particle
atomic
system are readily available by standard spectroscopic methods, which
require no access to the individual components. 
Given the large interest in spin squeezing,
a criterion of entanglement based on this property is an
important tool.
Recall, however, that systems may well be entangled without
being spin squeezed: The spin squeezing measurement provides a 
sufficient criterion for the depth of entanglement, not a necessary one.

We acknowledge support from the EU program EQUIP (contract
IST-1999-11053) and by Thomas B. Thriges Foundation.

\end{document}